\documentclass[a4paper]{jpconf}
\usepackage{graphicx}
\usepackage{citesort}
\begin{document}

\title{Parallelized Kalman-Filter-Based Reconstruction of Particle Tracks on Many-Core Architectures with the CMS Detector}

\author{G Cerati${}^4$, P Elmer${}^2$, B Gravelle${}^5$, M Kortelainen${}^4$, V Krutelyov${}^1$, S Lantz${}^3$,
M Masciovecchio${}^1$, K McDermott${}^3$, B Norris${}^5$, A Reinsvold Hall${}^4$, D Riley${}^3$, M Tadel${}^1$, P Wittich${}^3$, F W\"urthwein${}^1$ and A Yagil${}^1$}

\address{${}^1$ University of California, San Diego, La Jolla, CA, USA 92093}
\address{${}^2$ Princeton University, Princeton, NJ, USA 08544}
\address{${}^3$ Cornell University, Ithaca, NY, USA 14853}
\address{${}^4$ Fermilab, Batavia, IL, USA 60510-5011}
\address{${}^5$ University of Oregon, Eugene, OR, USA 97403}

\ead{mario.masciovecchio@cern.ch}

\begin{abstract}
In the High--Luminosity Large Hadron Collider (HL--LHC), one of the most challenging computational problems is expected to be finding and fitting charged-particle tracks during event reconstruction. The methods currently in use at the LHC are based on the Kalman filter. Such methods have shown to be robust and to provide good physics performance, both in the trigger and offline. In order to improve computational performance, we explored Kalman-filter-based methods for track finding and fitting, adapted for many-core SIMD and SIMT architectures. Our adapted Kalman-filter-based software has obtained significant parallel speedups using such processors, e.g., Intel Xeon Phi, Intel Xeon SP (Scalable Processors) and (to a limited degree) NVIDIA GPUs.
Recently, an effort has started towards the integration of our software into the CMS software framework, in view of its exploitation for the Run III of the LHC. Prior reports have shown that our software allows in fact for some significant improvements over the existing framework in terms of computational performance with comparable physics performance, even when applied to realistic detector configurations and event complexity. Here, we demonstrate that in such conditions physics performance can be further improved with respect to our prior reports, while retaining the improvements in computational performance, by making use of the knowledge of the detector and its geometry.
\end{abstract}

\section{Introduction}

Finding and fitting charged-particle tracks is one of the most computationally challenging steps of the event reconstruction in the CMS~\cite{cms} detector.
For the online reconstruction, it has a direct impact on the rate of data recorded by the CMS High Level Trigger (HLT), 
which are selected from the up to 100~kHz (up to 750~kHz in the High-Luminosity LHC, HL--LHC, era) 
CMS Level--1 trigger acceptance rate.
For the offline reconstruction, it is a limitation on the amount of data that can be processed for physics analyses.
The computational challenge represented by the track finding and fitting will become even more important in the HL--LHC era,
due to the increase in the instantaneous luminosity and in pileup (PU) of primary interactions per event, 
leading to an exponential increase in the track reconstruction time. 

With processors of all types gaining more and more of their performance from highly parallel elements,
the LHC experiments must exploit this feature in order to sustain the higher HL--LHC processing requirements.
We focus on the traditional Kalman Filter (KF) method~\cite{kf}, and adapt it to efficiently exploit highly parallel architectures,
such as Intel Xeon Phi, Intel Xeon SP (Scalable Processors), and NVIDIA GPGPUs.
Other parallel architectures have not been tested yet, as of today.
In this context, we have developed the {\sc{mkFit}} program, 
with the ultimate goal of reaching physics performance comparable with the current CMS track reconstruction~\cite{cmstrack}, 
while achieving a significant improvement in computational performance.

\section{Parallelized Kalman Filter tracking}

The {\sc{mkFit}} project started in 2014, with the development of a {\sc{Matriplex}} matrix operation library,
optimized for the simultaneous vectorized processing of sets of small matrices.
Relying on this library, the initial implementation of a vectorized KF track fitting algorithm was demonstrated in a simplified detector geometry~\cite{mkfit_fit},
followed by the initial implementation of an analogous track finding algorithm~\cite{mkfit_find}.
Further developments~\cite{mkfit_2016,mkfit_2017,mkfit_2018} allowed the software to achieve satisfactory performance,
using Intel Threading Building Blocks (TBB) parallel constructs to comply with the CMS software (CMSSW) code base.
A first implementation of the {\sc{mkFit}} program on GPGPUs was also pursued~\cite{mkfit_gpu}:
while the {\sc{Matriplex}} library was found to outperform standard small-matrix multiplication packages for GPUs, 
the performance of the KF-based track finding on GPGPUs is as of today not satisfactory.

After demonstrating satisfactory performance using a simplified detector geometry,
the {\sc{mkFit}} program was extended to handle a realistic detector geometry,
as well as to retain performance in realistic high detector occupancy scenarios, 
such as the high PU scenarios expected at the HL--LHC.

The latest developments, whose effects are illustrated in this document, 
aim at further improving the {\sc{mkFit}} physics performance
through the optimization of hit and track selection algorithms,
as well as at integrating the program with CMSSW, 
with the goal of integration in the CMS HLT test-bed system for Run~III of the LHC.
Future work will focus on the implementation of the Phase--II CMS geometry, 
with the ultimate goal of employing {\sc{mkFit}} in the CMS HLT, and possibly offline, 
during the HL--LHC era.

\section{The CMS detector geometry and events}

This document illustrates the performance of {\sc{mkFit}} 
for track reconstruction in simulated $t\bar{t}$ events with PU of 70 (50), using the Phase--I CMS detector geometry, 
that reflect the expected (actual) data taking conditions during the LHC Run~III (Run~II).
Two operation modes are available: standalone, where {\sc{mkFit}} operates independently of CMSSW; and within CMSSW.

\subsection{Geometry and detector description}
\label{sec:geom}

In the {\sc{mkFit}} program, geometry is described as a vector of \verb"LayerInfo" data structures containing:
physical dimensions of a layer; hit search windows; and auxiliary parameters and flags for track finding.
The last category includes information about layer detector type, 
individual detector module structure (double-sided, {\it{mono}}$+${\it{stereo}}, or single-sided, {\it{mono}}), 
and holes in detector coverage (currently only used for the CMS endcap detectors).

For track finding, {\it{tracking regions}} are defined together with the corresponding {\it{steering parameters}}.
In the current program, five distinct regions in $\eta$ are defined (barrel, $+z/-z$ transition, $+z/-z$ endcap).
Tracking regions could as well be defined according to track $p_{\mathrm{T}}$ 
in future development.
The steering parameters consist of a vector of \verb"LayerControl" data structures, containing indices of the layers 
to be traversed during track finding, as well as layer parameters and flags that are specific to a tracking region 
(e.g., information about potential seeding layers).
As a result, the track finding algorithm is agnostic of the actual detector structure, 
and it simply follows the layer-to-layer propagation defined by the steering parameters,
executing operations in accordance with the control flags contained in the \verb"LayerInfo" and \verb"LayerControl" data structures.

The setup described in this section is implemented as a plugin.
Hence, it is possible to support any detector geometry.
In the case of the Phase--I CMS geometry, effects of multiple scattering and energy loss are accounted for by defining 
two-dimensional arrays for radiation and interaction lengths, indexed in $r-z$.
These constants are taken from the CMS simulation.
The {\sc{mkFit}} program supports the usage of both constant and parametrized magnetic fields.

\subsection{Event handling and processing}
\label{sec:evt}

Hit and seed data are externally provided. 
In the standalone operation mode, the data are read from a binary file created by a dedicated converter application.
The same binary file can contain vectors of simulated tracks, and of tracks reconstructed by the standard CMS tracking, 
used to validate the physics performance of {\sc{mkFit}}.
When operating within CMSSW, {\sc{mkFit}} is used as an external software package, with a dedicated CMSSW processing module that is run within the CMSSW framework.
This module is responsible for packaging the input hit and seed data in the format expected by {\sc{mkFit}}, 
and for providing the configuration for the {\sc{mkFit}} execution. After the execution, found tracks are copied back into CMSSW format.
Data format conversions are currently undergoing optimization. 

Prior to the track finding, the seed collection undergoes a dedicated {\it{seed cleaning}} algorithm,
aimed at removing multiple instances of seeds that likely originate from hits belonging to the same outgoing particle.
This algorithm exploits the identity of hits, as well as the $p_{\mathrm{T}}$, $\eta$, and $\phi$ seed parameters,
and is tuned to preserve track finding efficiency in high PU events.
Duplicate seeds arise from detector module overlaps, which are especially significant in the endcaps of the CMS tracker.
Duplicate seeds are not removed prior to the standard CMS track finding, in CMSSW:
CMSSW processes seeds one by one, and when a track candidate is found its hits are marked as used;
if all the hits of a seed are marked as used from a previously found track candidate, that seed is not processed.
This approach is not practical as it would be a serial bottleneck in {\sc{mkFit}}, 
where up to $32\times N_{\mathrm{threads}}$ seeds are processed in parallel, 
and seeds are grouped to be close in $\eta$ and $\phi$ to maximize the memory cache reuse of hit data.

\section{Physics performance}

The physics performance of {\sc{mkFit}} is validated in simulated $t\bar{t}$ events with PU of 50 or 70, using the Phase--I CMS detector geometry.
A constant magnetic field of $3.8$~T is used.
Results correspond to the CMS {\it{initialStep}} tracking iteration, 
where seeds are required to have $4$ hits coming from distinct inner pixel layers, and to be compatible with the beam spot constraint~\cite{cmstrack}.
Equivalent results from the standard CMS tracking are also shown for comparison, using the same set of input seeds.

The results shown in this section must be considered as preliminary, 
since work is ongoing, as of the time of the conference, to finalize and optimize the {\sc{mkFit}} algorithm. 
For instance:
\begin{itemize}
\item hit search windows, track candidate scoring, and final track quality criteria are still undergoing optimization;
\item cleaning and merging of the final track collection is not yet implemented, which would include removal of duplicate tracks due to multiple seeds per simulated particle.
\end{itemize}

The physics performance of {\sc{mkFit}} is evaluated and validated using two validation suites:
\begin{itemize}
\item standalone validation: only used for the standalone {\sc{mkFit}} operation mode, 
  aimed at evaluating the {\sc{mkFit}} algorithm-level efficiency; 
\item CMSSW validation: aimed at evaluating and validating the absolute performance of {\sc{mkFit}}. 
\end{itemize}
In the standalone validation: reference tracks are simulated tracks with hits on at least 12 layers of the CMS tracker, including 4 layers from a seed track; 
a reconstructed track with $N_{\mathrm{hits}}\geq10$, with 4 of these hits belonging to a seed track, is considered matched to a reference track 
if at least 50\% of its hits belong to the latter, in addition to the 4 seed track hits.
In the CMSSW validation (as shown in this document): reference tracks are simulated tracks with $p_{\mathrm{T}}>0.9$~GeV, $|\eta|<2.5$, and $|d_{\mathrm{xy}}(\mathrm{vertex},\mathrm{beam~axis})|<3.5$~cm; 
a reconstructed track is considered matched to a reference track if $>75\%$ of its hits belong to the latter, 
with no additional selection 
and no seed hit requirement.
In both cases, track reconstruction efficiency is defined as the fraction of reference tracks with (at least) a matching reconstructed (found) track.

\begin{figure}[h]
\begin{minipage}{0.45\linewidth}
\includegraphics[width=\linewidth]{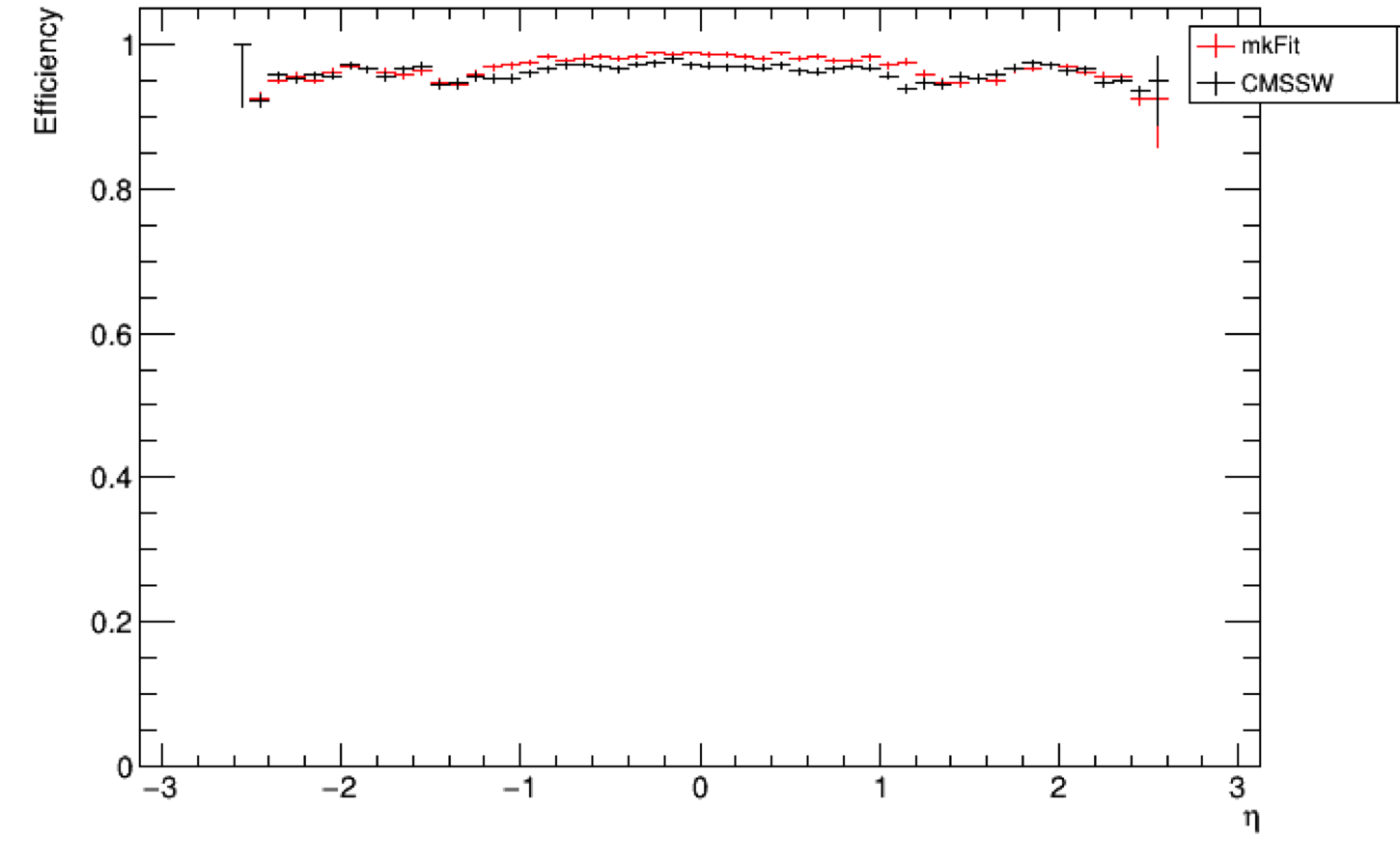}
\caption{\label{fig:mkfiteff}Track reconstruction efficiency, as obtained from the {\sc{mkFit}} standalone validation suite, 
as a function of the track 
$\eta$ for tracks with $p_{\mathrm{T}}>0.9$~GeV.
The threshold $p_{\mathrm{T}}>0.9$~GeV corresponds to the target minimum $p_{\mathrm{T}}$ threshold for CMS HLT operation.}
\end{minipage}\hspace{2pc}%
\begin{minipage}{0.45\linewidth}
\includegraphics[width=\linewidth]{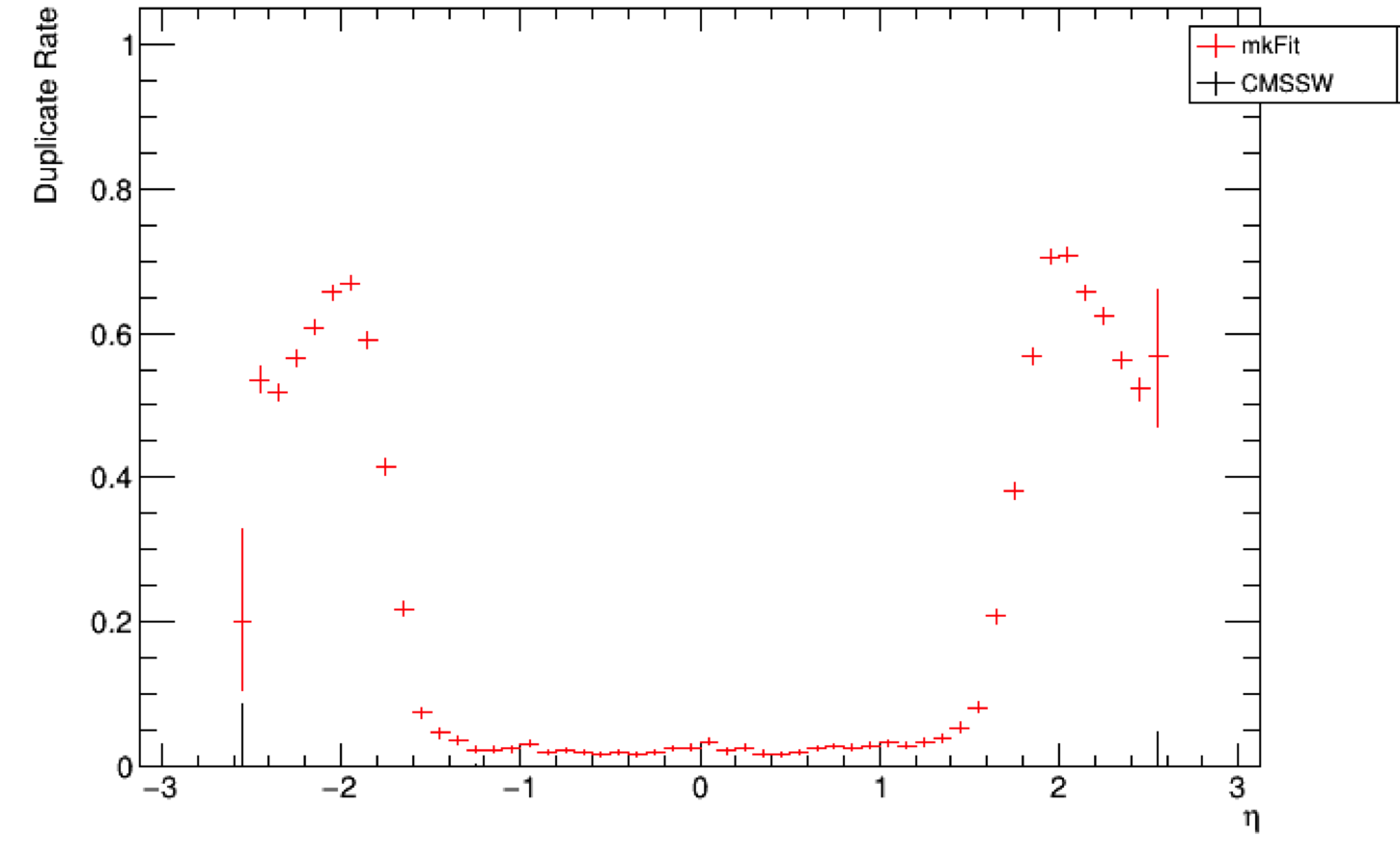}
\caption{\label{fig:mkfitdr}Duplicate track rate, as obtained from the {\sc{mkFit}} standalone validation suite, 
as a function of the track 
$\eta$ for tracks with $p_{\mathrm{T}}>0.9$~GeV.
The threshold $p_{\mathrm{T}}>0.9$~GeV corresponds to the target minimum $p_{\mathrm{T}}$ threshold for CMS HLT operation.
CMSSW values are nearly zero.}
\end{minipage}
\end{figure}

Figure~\ref{fig:mkfiteff} shows the track reconstruction efficiency as obtained from the standalone validation suite, for both {\sc{mkFit}} and CMSSW tracks, 
as a function of the track 
$\eta$, in simulated $t\bar{t}$ events with PU of 70:
the {\sc{mkFit}} algorithm-level efficiency is as good as the standard CMSSW track reconstruction, or larger.
The duplicate track rate as obtained from the standalone validation suite in the same simulated events, for both {\sc{mkFit}} and CMSSW tracks, 
is shown in Fig.~\ref{fig:mkfitdr} as a function of the track 
$\eta$: the {\sc{mkFit}} duplicate rate is especially large for endcap tracks,
mainly due to module overlaps in this region of the CMS tracker, and to the absence, in {\sc{mkFit}}, of a dedicated duplicate track removal procedure.
As of the time of the conference presentation, an effort is ongoing for the implementation of such a procedure.
%

\begin{figure}[h]
\begin{minipage}{0.45\linewidth}
\vspace{1.mm}
\includegraphics[height=1.0\linewidth]{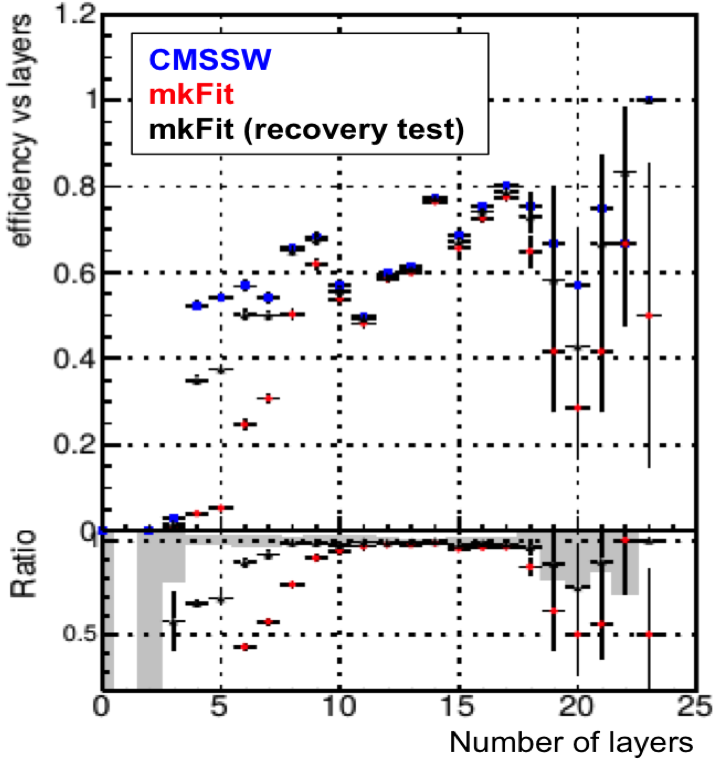}
\caption{\label{fig:cmssweff}Track reconstruction efficiency in simulated $t\bar{t}$ events with PU of 50, as obtained from the CMSSW validation suite, 
as a function of $N_{\mathrm{layers}}$, for CMSSW tracks, {\sc{mkFit}} tracks, 
and {\sc{mkFit}} tracks where a test was performed to (partially) recover the inefficiency otherwise observed for tracks with $N_{\mathrm{layers}}<12$.}
\end{minipage}\hspace{2pc}%
\begin{minipage}{0.45\linewidth}
\includegraphics[height=1.0\linewidth]{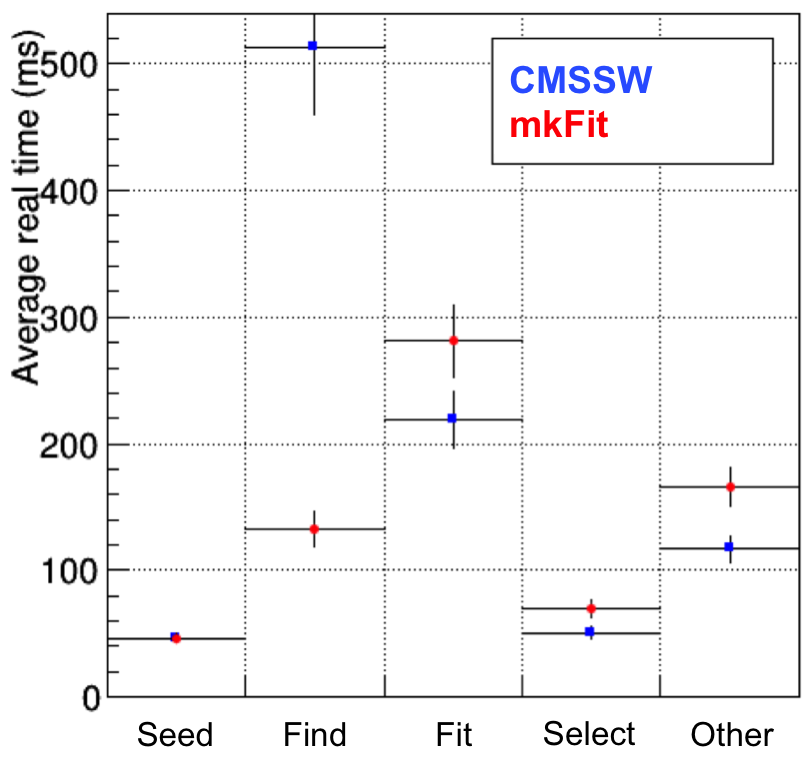}
\caption{\label{fig:cmsswtime}Average track reconstruction time, 
for both {\sc{mkFit}} and CMSSW 
tracks, when {\sc{mkFit}} is operated within CMSSW, in simulated $t\bar{t}$ events with PU of 50, on 
SKL--SP, Skylake Gold, Intel Xeon Gold 6130 CPU @ 2.1~GHz,
using a single thread. Each bin is a step of the standard CMSSW track reconstruction.} 
\end{minipage}
\end{figure}

Figure~\ref{fig:cmssweff} shows the track reconstruction efficiency as obtained from the CMSSW validation suite, for both {\sc{mkFit}} and CMSSW tracks, 
as a function of the number of tracker layers, in simulated $t\bar{t}$ events with PU of 50, when {\sc{mkFit}} is operated within CMSSW: 
a relative inefficiency of {\sc{mkFit}} is observed, compared to CMSSW tracks, for tracks with $N_{\mathrm{layers}}<12$.
The inefficiency is related to partial {\sc{mkFit}} algorithm optimizations based on the {\sc{mkFit}} standalone validation suite.
Work is ongoing to recover the lost efficiency, as of the time of the conference presentation. 
The preliminary results of a test performed for this purpose are shown in Fig.~\ref{fig:cmssweff}.

\section{Computational performance}

The computational performance is evaluated on the most recent development platform: 
SKL--SP, Skylake Gold, $2~\mathrm{sockets}\times16~\mathrm{cores}$ with hyperthreading enabled, Intel Xeon Gold 6130 CPU @ 2.1~GHz,
with the Turbo Boost feature disabled.
The Intel \verb"icc" compiler is used, together with the \verb"AVX-512" set of instructions.

The track reconstruction performance in simulated $t\bar{t}$ events with PU of 50, using a single thread,
is shown in Fig.~\ref{fig:cmsswtime}, where {\sc{mkFit}} is compared to the standard CMS track reconstruction.

For track finding, {\sc{mkFit}} is found to perform better than CMSSW, by a factor of about $4$.
This factor also accounts for the data format conversions described in Sec.~\ref{sec:evt},
which make up about 40\% of the total {\sc{mkFit}} track finding time.
Once the data conversion operations have been optimized, {\sc{mkFit}} can be expected 
to achieve even better performance compared to CMSSW,
by up to a factor of about $7$.
The performance advantage of {\sc{mkFit}}, as compared to CMSSW standard tracking, 
is understood as coming from two sources: vectorization, and different choices with respect to CMSSW 
for the implementation of geometry and detector description (see Sec.~\ref{sec:geom}).  
%
%

For both the {\sc{mkFit}} and CMSSW results, final track fitting is performed within CMSSW using its standard fitting procedure.
In addition to the track KF fit, this procedure includes hit outlier rejection, based on a precise cluster position estimate 
for the pixel layers of the detector.
Interestingly, the final track fitting takes longer for {\sc{mkFit}}. 
This can be attributed to the larger duplicate track rate observed for {\sc{mkFit}}.
The in-progress (as of the time of the conference presentation) {\sc{mkFit}} duplicate track removal procedure is expected 
to cancel the currently observed deficit of performance.

\begin{figure}[h]
\includegraphics[width=0.5\linewidth]{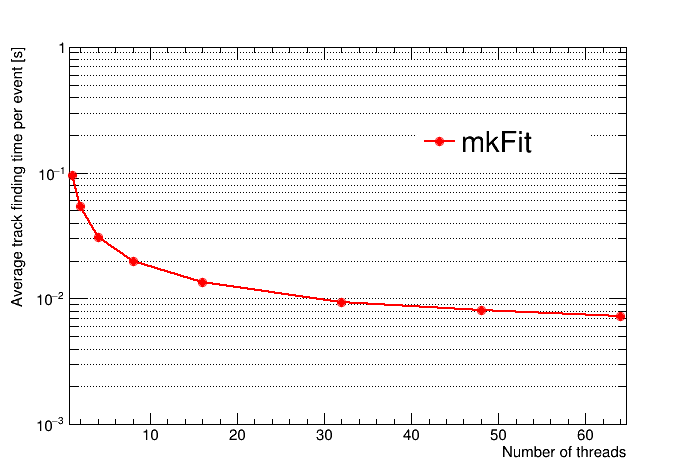}
\includegraphics[width=0.5\linewidth]{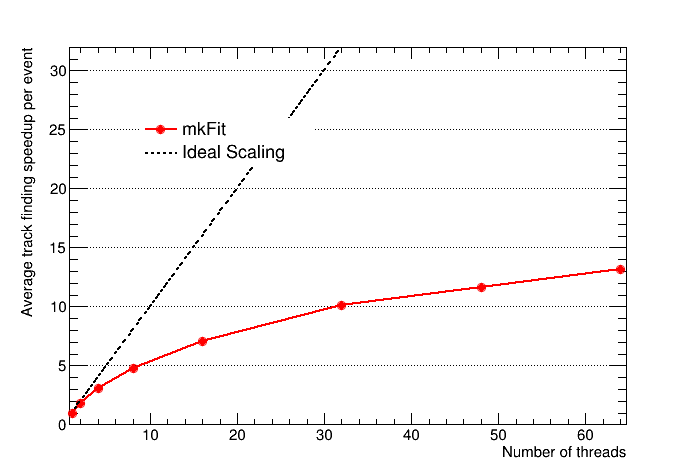}
\caption{\label{fig:mkfittime}Average {\sc{mkFit}} track finding time per event (left) and corresponding speedup (right) as a function of the used number of threads, 
on SKL--SP, in simulated $t\bar{t}$ events with PU of 70, using the standalone {\sc{mkFit}} operation mode.}
\end{figure}

The {\sc{mkFit}} algorithm is also able to exploit multithreading, and when this is enabled within each event, 
the per-event timing is demonstrated to scale with the number of threads as shown in Fig.~\ref{fig:mkfittime}.
In this case, track finding is performed on simulated $t\bar{t}$ events with PU of 70, using the standalone {\sc{mkFit}} operation mode,
on the same SKL--SP architecture.

\section{Conclusion and outlook}

An algorithm for parallelized Kalman-Filter-based reconstruction of charged-particle tracks on multi-/many-core architectures has been developed, 
and its performance has been demonstrated using the CMS detector at the LHC, with conditions that reflect what is expected during the LHC Run~III.
The preliminary results summarized in this document show that {\sc{mkFit}} can achieve comparable physics performance with respect to traditional KF tracking algorithms,
while retaining a significant improvement in computational performance.
Ongoing work focuses on the optimization of the track finding algorithm parameters, as well as on the implementation of the final track post-processing steps.

Integration of {\sc{mkFit}} with CMSSW is also advancing, on a path for a full integration in the CMS HLT test-bed system for the LHC Run~III.
In this context, work is ongoing on the optimization of data format conversion operations.
Future work will focus on the implementation of the Phase--II CMS detector, 
with the ultimate goal of employing {\sc{mkFit}} in the CMS HLT, and possibly offline, 
during the HL--LHC era.

\section*{Acknowledgments}
This work is supported by the U.S. National Science Foundation, under grants PHY--1520969, PHY--1521042, PHY--1520942 and PHY--1624356, and under Cooperative Agreement OAC-1836650,
and by the U.S. Department of Energy, Office of Science, Office of Advanced Scientific Computing Research, Scientific Discovery through Advanced Computing (SciDAC) program.

\end{document}